\documentclass{ws-ijmpe}

\begin{document}

\markboth
{M. G\'o\'zd\'z and W. A. Kami\'nski} 
{$0\nu2\beta$ Nuclear Matrix Elements and Neutrino Magnetic Moments}

%
%

\title{$0\nu2\beta$ NUCLEAR MATRIX ELEMENTS \\ 
  AND NEUTRINO MAGNETIC MOMENTS}

\author{MAREK G\'O\'ZD\'Z and WIES{\L}AW A. KAMI\'NSKI}

\address{
Department of Informatics, Maria Curie-Sk{\l}odowska University, \\
pl. Marii Curie--Sk{\l}odowskiej 5, 20-031 Lublin, Poland \\
mgozdz@kft.umcs.lublin.pl, kaminski@neuron.umcs.lublin.pl
}

\maketitle

\begin{history}                 %
\end{history}                  	%

\begin{abstract}
  We compare different methods of obtaining the neutrinoless double beta
  decay nuclear matrix elements (NME). On the example of ${}^{76}$Ge we
  use the NME to calculate the Majorana neutrino transition magnetic
  moments, generated through particle--sparticle $R$-parity violating
  loop diagrams whithin the minimal supersymmetric standard model.
\end{abstract}

\section{Introduction}

The neutrinoless double beta ($0\nu2\beta$) decay is a~hypothetical
nuclear process, in which two simultaneous beta decays occur in one
nucleus, whereas some exotic mechanism prevents the neutrinos to be
emitted. It would result of course in the lepton number violation
($\Delta L=2$) and therefore it is forbidden by the standard
theory. However, in certain exotic models of physics beyond the standard
model, such processes are allowed and sought in a~number of
experiments. The observation of a~$0\nu2\beta$ decay will open
a~completely new area of physics, which by far is only a~speculation.

The theory of this exotic decay involves two parts: firstly, as this is
a~nuclear process it requires a~careful calculation of the nuclear
matrix elements; secondly, an exotic mechanism must be proposed, which
would suppress the emission of the neutrinos. The $0\nu2\beta$ half-life
may therefore be expressed in a~factorized form as
\begin{equation}
  ( T^{0\nu} )^{-1} = G^{0\nu} |M^{0\nu}|^2 |\langle m_{ee} \rangle|^2,
\end{equation}
where $G^{0\nu}$ is the exactly calculable phase-space factor,
$M^{0\nu}$ is the NME, and $\langle m_{ee} \rangle$ is the so-called
effective neutrino mass, which represents the non-standard interactions
involved. We are not going to discuss the latter part here.

In this communication, we are going to compare different methods of
obtaining the nuclear matrix elements (NME) for the $0\nu2\beta$
decay. As an application, we will use them to calculate the Majorana
neutrino transition magnetic moments within the minimal supersymmetric
standard model with broken $R$-parity.

\section{Calculations of the NME: random phase approximation}

The random phase approximation (RPA) methods in their simplest form base
on the assumption, that the ground state is expressed as a~BCS
vacuum. Then, a~transition amplitude between the $0^+$ BCS vacuum of an
even--even nucleus and excited states of the neighboring odd--odd
nucleus is calculated, representing the transition between the mother
nucleus and an intermediate one. A~similar amplitude represents the
transition between the intermediate and the daughter nucleus. These two
parts are then summed over possible excited states of the intermediate
virtual nucleus, which are expressed as simple harmonic oscillations
above its BCS ground state.

Such simplified picture does not give satisfying results and many
variation of the original method appear. These involve the use of
quasiparticles (QRPA, pnQRPA), renormalized QRPA (RQRPA), selfconsistent
RQRPA (SRQRPA) and others. In QRPA a quasiboson approximation is used,
in which the commutation relations for fermions are replaced by bosonic
ones, and this procedure seems to be working well for small harmonic
excitations, but breaks down when one wants to take into account higher
excitations as well. In RQRPA the commutators are approximated in a
different way, which introduces the new factor as a normalization
constant.

What is more, different schemes of short-range correlations are used,
among which the Jastrow and the unitary correlation operator method
(UCOM) are most common. Also one may adjust other parameters, such as
the nuclear potential (Woods--Saxon, effective mean field etc.), and the
strength of the particle-particle ($g_{pp}$) and particle-hole ($g_{ph}$)
interactions. 

All these give a~quite big diversity in the outcome of different
calculations. At present one may summarize the results of the
Jyv\"askyl\"a and the T\"ubingen groups,\cite{mgozdz:qrpa} for $^{76}$Ge
as $3.33 < M^{0\nu} < 6.64$.

\section{Calculations of the NME: nuclear shell model}

The nuclear shell model (NSM) and its recently announced large--scale
version (LSSM) is meant in principle to give exact results. The approach
is straightforward and involves defining a~valence space, deriving an
effective interaction from the Hamiltonian, building, and then finally
diagonalizing the Hamiltonian matrix. The first and most obvious
obstacle here is, however, the dimension of the matrix in question,
which is proportional to
\begin{equation}
  \mathrm{dim} \sim 
  \left(\begin{array}{c} d_\pi \\ p \end{array}\right)
  \cdot 
  \left(\begin{array}{c} d_\nu \\ n \end{array}\right)
\end{equation}
where $d_\pi$ ($d_\nu$) is the dimension of the proton (neutron)
subshell, and $p$ ($n$) is the number of valence protons
(neutrons). This feature makes it practically impossible to include in
the valence space all important single particle orbits; eg. the
spin-orbit partners are usualy neglected. Also in the actual
calculations the many-body problem is reduced to two-body problem within
some mean-field approach.

The most advanced large scale shell model calculations for the $^{76}$Ge
nucleus yield at present $M^{0\nu}=2.22$ (without higher order
contributions to the nuclear current this value increases to
2.58). A~little bit older calculations which used the UCOM short-range
correlations gave $M^{0\nu}=2.81$.\cite{mgozdz:LSSM}

\section{Calculations of the NME: interacting boson model}

The interacting boson model (IBM) has not been used for this type of
calculations before, but it turned out to give results which
surprisingly well agree with the (R)QRPA calculations. In this method,
a~fermionic two-body matrix element is used to obtain the fermions
transition operator in second quantization. The matrix elements of this
operator are then evaluated in the general seniority scheme. At the end,
the fermionic operator is mapped into bosonic transition operator, whose
matrix elements are evaluated using realistic IBM wave function.

This method, introduced by Barea and Iachello in Ref.\cite{mgozdz:IBM},
yields for the $^{76}$Ge nucleus $4.64 < M^{0\nu} < 5.46$, which is in
very good agreement with the central values obtained within the (R)QRPA
method.

\begin{table}
  \tbl{\label{mgozdz:tab1}Neutrinoless double beta decay nuclear matrix
    element $M^{0\nu}$ for $^{76}$Ge calculated using three different
    approaches. The last column shows the corresponding upper limit on
    the effective neutrino mass (see text for details).}{
\begin{tabular}{lccc}
  \toprule
  method & $M^{0\nu}$: ranges & central value & $|\langle m_{ee} \rangle| \le$\\
  \colrule
  (R)QRPA & 3.33 -- 6.64 & 4.985 $\pm$ 1.655 & 0.22 -- 0.43 eV \\
  LSSM    & 2.22 -- 2.81 & 2.515 $\pm$ 0.295 & 0.51 -- 0.65 eV \\
  IBM     & 4.64 -- 5.46 & 5.050 $\pm$ 0.414 & 0.26 -- 0.31 eV \\
  \botrule
\end{tabular}}
\end{table}

We summarize the various results for the $^{76}$Ge nucleus in
Tab.\ref{mgozdz:tab1}. In the last column we have calculated the
effective neutrino mass using the result of the Heidelberg--Moscow
experiment, $T_{1/2}^{0\nu}(^{76}\hbox{Ge}) \le 1.9 \times 10^{25}
\hbox{ y}$,\cite{mgozdz:HM} and a~common phase-space factor
$G^{0\nu}=2.55 \times 10^{-26}$ y eV$^2$.

\section{Majorana neutrino transition magnetic moments in $R$-parity
  violating supersymmetry}

\begin{figure}
  \hbox{\vbox{
  \includegraphics[width=0.30\textwidth]{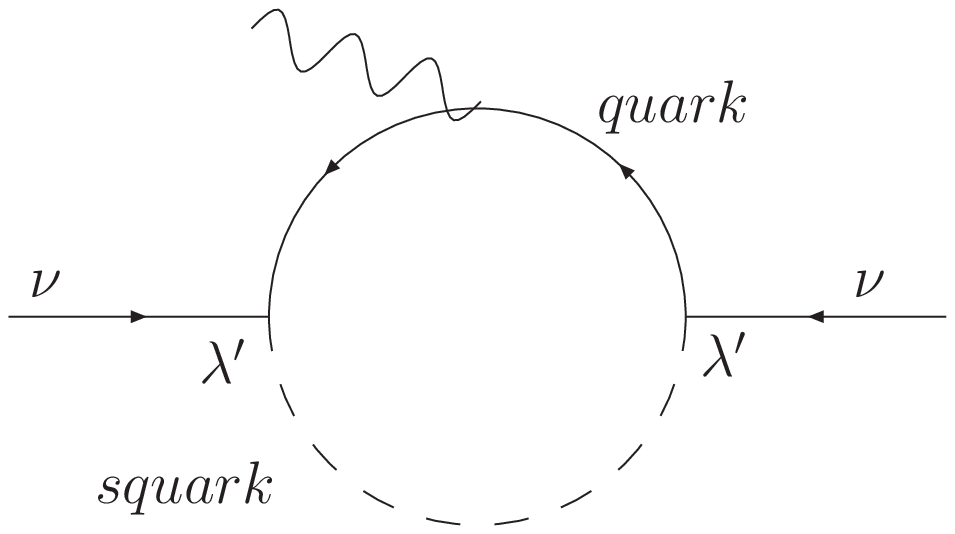}
  \includegraphics[width=0.30\textwidth]{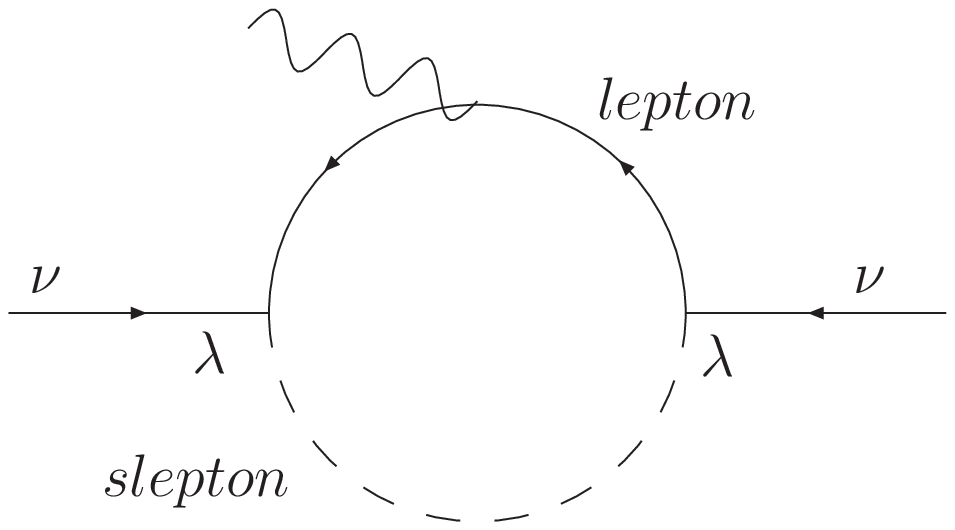} \\
  \includegraphics[width=0.30\textwidth]{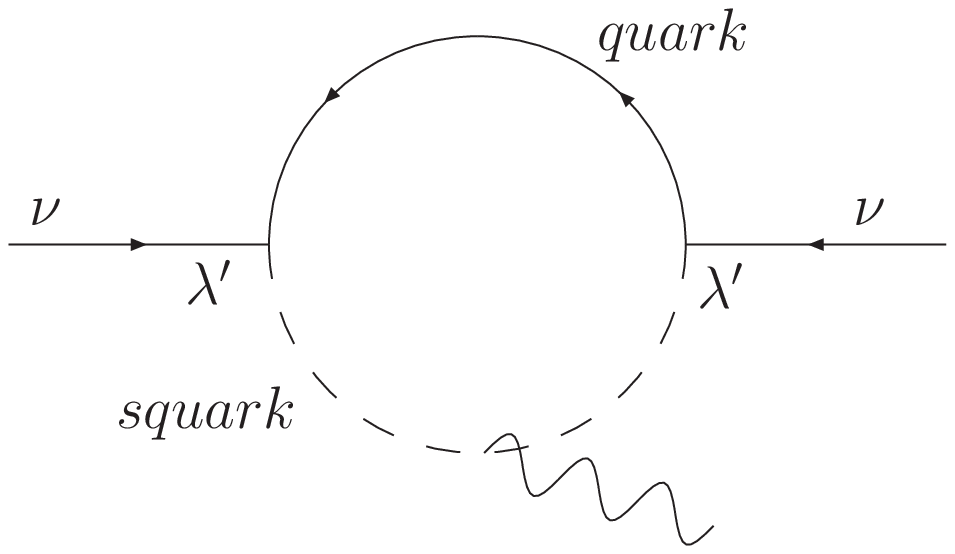}
  \includegraphics[width=0.30\textwidth]{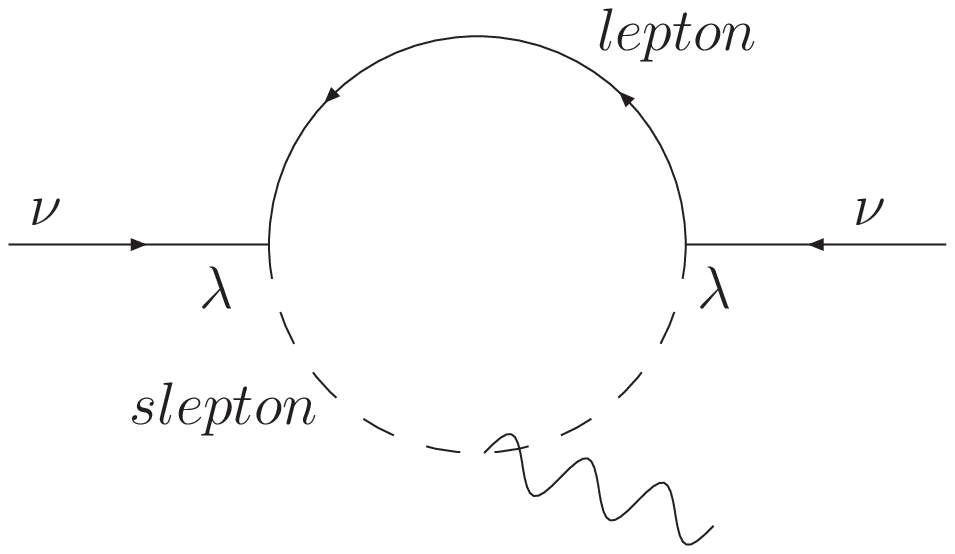}}
  \hskip -4.5 truecm
  \vbox{
    \begin{minipage}{0.33\textwidth}
      \vspace{-7truecm}
  \caption{\label{mgozdz:fig-diag} Processes leading to Majorana
    neutrino transition magnetic moments.}
\end{minipage}
}}
\end{figure}

The Majorana magnetic moment acts between $\nu_{i L}$ and $\nu_{j L}^c$
chiral components of Majorana neutrinos, assuming a~standard gauge
theory with only left-handed neutrinos. As a consequence, it violates
the total lepton number by two units $\Delta L =2$. The effective
Hamiltonian $H_{\rm eff}$ of this interaction takes the form
\begin{equation}
  H_{\rm eff} = \frac{1}{2} \mu_{ij} \bar\nu_{iL} \sigma^{\alpha\beta}
  \nu_{j L}^c F_{\beta\alpha} + {\rm h.c.},
\end{equation}
where $F$ is the electromagnetic field strength tensor, and
$\sigma^{\alpha\beta}$ is defined by means of the Pauli matrices,
$\sigma^{\alpha\beta}=\frac12 [\sigma^\alpha,\sigma^\beta]$. We work in
the minimal supersymmetric standard model with broken
$R$-parity.\cite{mgozdz:MSSM} In this setting it is possible to consider
processes, in which the Majorana neutrino, despite being electrically
neutral, may effectively interact with an external photon. The amplitude
of such processes is governed by the magnetic moments of the
neutrino. We consider interaction between two Majorana neutrinos, in
which the effective interaction vertex is expanded into
a~particle--sparticle loop, as depicted on Fig.\ref{mgozdz:fig-diag}

The quark--squark loop contribution, including possible $d$-quark mixing
($V$ is the CKM matrix), is given by the following
formula:\cite{mgozdz:mg}
\begin{equation}
  \mu_{\nu_{ii'}}^q = (1-\delta_{ii'}) \frac{12 Q_d m_e}{16\pi^2}
  \sum_{jkl} \left \{
  \lambda'_{ijk}\lambda'_{i'kl} \sum_a V_{ja} V_{la}
  \frac{w^q_{ak}}{m_{d^a}}
   -  (k \leftrightarrow j) \right \} \mu_B,
\end{equation}
where $w$ is the loop integral, which arises from the integration over
virtual momenta of particles inside the loop,
\begin{equation}
  w^q_{jk} = \frac{\sin 2\theta^k}{2} \left (
    \frac{x_2^{jk}\ln x_2^{jk}-x_2^{jk}+1}{(1-x_2^{jk})^2} -
    (x_2 \to x_1) \right ),
\end{equation}
and we have denoted the ratio of the quark and squark masses squared by
$x_i^{jk} = m_{d^j}^2 / m_{\tilde d_i^k}^2$. Here $\theta^k$ is the
$k$-th squark mixing angle. The contribution from the slepton--lepton
loop does not contain summation over three quark colors, and we do not
include the very weak mixing in the charged lepton sector. The formula
reads therefore\cite{mgozdz:mg}
\begin{equation}
  \mu_{\nu_{ii'}}^\ell = (1-\delta_{ii'}) \frac{4 Q_e m_e}{16\pi^2}
  \sum_{jk}
  \lambda_{ijk}\lambda_{i'kj} \left(
    \frac{w^\ell_{jk}}{m_{e^j}} - \frac{w^\ell_{kj}}{m_{e^k}} \right ) 
  \mu_B,
\end{equation}
where the loop integral is identical as previously with the exception
that $x_i^{jk} \to y_i^{jk} = m_{\ell^j}^2 / m_{\tilde \ell_i^k}^2$, and
$\theta^k \to \phi^k$, the slepton mixing angle.

\begin{figure}[t]
  \hbox{\vbox{
    \includegraphics[width=0.45\textwidth]{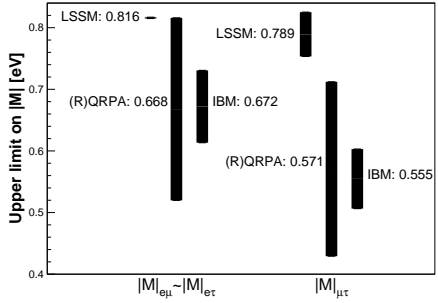}}
  \hskip -6.5 truecm
  \vbox{
    \begin{minipage}{0.5\textwidth}
      \vspace{-6.5 truecm}
      \caption{\label{mgozdz:fig-comparison}Ranges for the maximal values of
        the off-diagonal elements in the neutrino mass matrix, coming from
        different $0\nu2\beta$ nuclear matrix elements.}
  \end{minipage}
}}
\end{figure}


In order to get numerical results, we constrain the model by imposing
grand unification at very high energy scale, $M_{\rm GUT} \approx 1.2
\times 10^{16}$ GeV, thus starting from few free parameters only. These
are the common mass of scalars $m_0$, the common mass of fermions
$m_{1/2}$, common Yukawa coupling constants unification factor $A_0$,
the ratio of the Higgs vacuum expectation values $\tan\beta$, and the
sign of the bilinear up- and down-type Higgs coupling sgn$(\mu)$. Next,
the renormalization group equations are used to derive the values of the
couplings and mass parameters at low energy scales. The $R$-parity
violating trilinear couplings $\lambda$ and $\lambda'$, which cannot be
derived from the GUT constraints, are assessed from the experimental
neutrino mass matrix. This procedure allows to calculate the transition
magnetic moments (see Ref.\cite{mgozdz:mg} for details).

\begin{table}
  \tbl{\label{mgozdz:tab2}Majorana neutrino transition magnetic moments
    $\mu_{\nu_{ij}}$ in $\mu_B$ for GUT parameters: $A_0=100$ GeV,
    $m_0=m_{1/2}=150$ GeV, $\tan\beta=19$, $\mu>0$. Ranges correspond to
    the spread in NME obtained using different methods.}{
\begin{tabular}{cccc}
  \toprule
  $ij$ & LSSM & (R)QRPA & IBM \\ 
  \colrule
  & \multicolumn{3}{c}{lepton-slepton loop mechanism} \\
  $e\mu, e\tau$ & $(1.33, 1.33)\ 10^{-15}$ & $(8.46, 13.3)\ 10^{-16}$ & $(9.99, 11.9)\ 10^{-16}$ \\
  $\mu\tau$     & $(1.23, 1.35)\ 10^{-15}$ & $(7.02, 11.6)\ 10^{-16}$ & $(8.28, 9.86)\ 10^{-16}$ \\\\
  & \multicolumn{3}{c}{quark-squark loop mechanism (without d-quarks mixing)} \\
  $e\mu, e\tau$ & $(9.49, 9.49)\ 10^{-17}$ & $(6.04, 9.49)\ 10^{-17}$ & $(7.13, 8.50)\ 10^{-17}$ \\ 
  $\mu\tau$     & $(8.69, 9.52)\ 10^{-17}$ & $(4.95, 8.22)\ 10^{-17}$ & $(5.84, 6.96)\ 10^{-17}$ \\\\
  & \multicolumn{3}{c}{quark-squark loop mechanism (with d-quarks mixing)} \\
  $e\mu, e\tau$ & $(8.22, 8.22)\ 10^{-17}$ & $(5.24, 8.22)\ 10^{-17}$ & $(6.18, 7.36)\ 10^{-17}$ \\
  $\mu\tau$     & $(7.24, 7.93)\ 10^{-17}$ & $(4.13, 6.85)\ 10^{-17}$ &
  $(4.87, 5.80)\ 10^{-17}$ \\
  \botrule
\end{tabular}}
\end{table}

\begin{table}[h]
  \tbl{\label{mgozdz:tab3}Majorana neutrino transition magnetic moments
    $\mu_{\nu_{ij}}$ in $\mu_B$ for GUT parameters: $A_0=500$ GeV,
    $m_0=m_{1/2}=1000$ GeV, $\tan\beta=19$, $\mu>0$.  Ranges correspond
    to the spread in NME obtained using different methods.}{
\begin{tabular}{cccc}
  \toprule
  $ij$ & LSSM & (R)QRPA & IBM \\ 
  \colrule
  & \multicolumn{3}{c}{lepton-slepton loop mechanism} \\
  $e\mu, e\tau$ & $(3.48, 3.48)\ 10^{-17}$ & $(2.22, 3.48)\ 10^{-17}$ & $(2.62, 3.12)\ 10^{-17}$ \\
  $\mu\tau$     & $(3.26, 3.57)\ 10^{-17}$ & $(1.86, 3.08)\ 10^{-17}$ & $(2.19, 2.61)\ 10^{-17}$ \\\\
  & \multicolumn{3}{c}{quark-squark loop mechanism (without d-quarks mixing)} \\
  $e\mu, e\tau$ & $(2.53, 2.53)\ 10^{-18}$ & $(1.61, 2.53)\ 10^{-18}$ & $(1.90, 2.27)\ 10^{-18}$ \\ 
  $\mu\tau$     & $(2.32, 2.54)\ 10^{-18}$ & $(1.32, 2.19)\ 10^{-18}$ & $(1.56, 1.86)\ 10^{-18}$ \\\\
  & \multicolumn{3}{c}{quark-squark loop mechanism (with d-quarks mixing)} \\
  $e\mu, e\tau$ & $(2.15, 2.15)\ 10^{-18}$ & $(1.37, 2.15)\ 10^{-18}$ & $(1.62, 1.93)\ 10^{-18}$ \\  
  $\mu\tau$     & $(2.03, 2.22)\ 10^{-18}$ & $(1.16, 1.92)\ 10^{-18}$ &
  $(1.36, 1.62)\ 10^{-18}$ \\
  \botrule
\end{tabular}}
\end{table}


Fig.\ref{mgozdz:fig-comparison} presents how the off-diagonal elements
in the neutrino mass matrix depend on the nuclear matrix element
used. Each bar shows the range for a~given method, with the number being
its central value. One sees immediately, that the (R)QRPA and IBM
methods agree very well with each other, while the shell model
calculations give substantially higher results. This trend is of course
preserved also in the magnetic moment data, which are shown in
Tabs.\ref{mgozdz:tab2} and \ref{mgozdz:tab3}. We have included
calculations for two cases, one with `small' GUT scale unification
parameters, the other with `large' ones. For each case three
possibilities were considered, namely the lepton--slepton loop, and the
quark--squark loop with or without additional $d$-quark mixing. 
One sees that the differences in the magnetic moments for various
nuclear matrix elements are negligible, and that the ranges often
overlap with each other. As expected, the LSSM gives the highest
results, but the difference between for example the upper (R)QRPA value
and the lower LSSM one is roughly of the order of $10\%$. Of course, the
predictions on the half-life of the $0\nu2\beta$ decay, or the effective
neutrino mass may differ in a~much more serious manner. In such a~case
the (R)QRPA and IBM methods are worth recommendation.

\noindent {\bf Acknowledgements. } 
The first author (MG) greatly acknowledges financial support from the
Polish State Committee for Scientific Research under grant no.
N~N202~0764~33.



\end{document}